\documentclass[twocolumn,superscriptaddress,showpacs]{revtex4-2}
\usepackage{graphicx}%
\usepackage{dcolumn}
\usepackage{amsmath}
\usepackage{amssymb}
\usepackage{bm}
\usepackage{color}
\usepackage[whole]{bxcjkjatype}

\begin{document}

%\preprint{HEP/123-qed}

%%%%%%%%%%%%%%%%%*Title*%%%%%%%%%%%%%%%%%%%

\preprint{submitted to Physical Review B}
\title{
Bulk superconductivity in Pb-substituted BiS$_{\bf 2}$-based compounds
studied by hard-x-ray spectroscopy
}

\author{A. Yamasaki}
\email[Author to whom correspondence should be addressed. \\
E-mail: ]{yamasaki@konan-u.ac.jp}
\affiliation{Faculty of Science and Engineering, Konan University, Kobe 658-8501, Japan }
\affiliation{RIKEN SPring-8 Center, Sayo, Hyogo 679-5148, Japan}

\author{T. Oguni}
\affiliation{Graduate School of Natural Science, Konan University, Kobe 658-8501, Japan }

\author{T. Hayashida}
\author{K. Miyazaki}
\author{N. Tanaka}
\author{K.~Nakagawa}
\affiliation{RIKEN SPring-8 Center, Sayo, Hyogo 679-5148, Japan}
\affiliation{Graduate School of Natural Science, Konan University, Kobe 658-8501, Japan }

\author{K. Tamura}
\affiliation{Graduate School of Engineering, Osaka Prefecture University, Sakai, Osaka 599-8531, Japan}

\author{K.~Mimura}
\affiliation{Graduate School of Engineering, Osaka Prefecture University, Sakai, Osaka 599-8531, Japan}
\affiliation{Graduate School of Engineering, Osaka Metropolitan University, Sakai, Osaka 599-8531, Japan}

\author{N. Kawamura}
\affiliation{Japan Synchrotron Radiation Research Institute, Sayo, Hyogo 679-5198, Japan}

\author{H. Fujiwara}
\affiliation{RIKEN SPring-8 Center, Sayo, Hyogo 679-5148, Japan}
\affiliation{Graduate School of Engineering Science, Osaka University, Toyonaka, Osaka 560-8531, Japan}
\affiliation{
Spintronics Research Network Division, 
Institute for Open and Transdisciplinary Research Initiatives,
Osaka University, 
Suita, Osaka 565-0871, Japan}
\author{G.~Nozue}
\author{A. Ose}
\affiliation{RIKEN SPring-8 Center, Sayo, Hyogo 679-5148, Japan}
\affiliation{Graduate School of Engineering Science, Osaka University, Toyonaka, Osaka 560-8531, Japan}

\author{Y. Kanai-Nakata}
\affiliation{RIKEN SPring-8 Center, Sayo, Hyogo 679-5148, Japan}
\affiliation{College of Science and Engineering, Ritsumeikan University, Kusatsu, Shiga 525-8577, Japan}

\author{A. Higashiya}
\affiliation{RIKEN SPring-8 Center, Sayo, Hyogo 679-5148, Japan}
\affiliation{Faculty of Science and Engineering, Setsunan University, Neyagawa, Osaka 572-8508, Japan}

\author{S.~Hamamoto}
\author{K. Tamasaku}
\author{M.~Yabashi}
\author{T. Ishikawa}
\affiliation{RIKEN SPring-8 Center, Sayo, Hyogo 679-5148, Japan}

\author{S. Imada}
\affiliation{RIKEN SPring-8 Center, Sayo, Hyogo 679-5148, Japan}
\affiliation{College of Science and Engineering, Ritsumeikan University, Kusatsu, Shiga 525-8577, Japan}

\author{A. Sekiyama}
\affiliation{RIKEN SPring-8 Center, Sayo, Hyogo 679-5148, Japan}
\affiliation{Graduate School of Engineering Science, Osaka University, Toyonaka, Osaka 560-8531, Japan}
\affiliation{
Spintronics Research Network Division, 
Institute for Open and Transdisciplinary Research Initiatives,
Osaka University, 
Suita, Osaka 565-0871, Japan}

\author{H. Sakata}
\affiliation{Department of Physics, Tokyo University of Science, Tokyo 162-8601, Japan}

\author{S. Demura}
\affiliation{College of Science and Technology, Nihon University, Tokyo 101-8308, Japan}

\date{\today}

\begin{abstract}

In this study, we investigate the bulk electronic structure of Pb-substituted LaO$_{0.5}$F$_{0.5}$BiS$_2$ 
single crystals, using two types of hard-x-ray spectroscopy. 
High-energy-resolution fluorescence-detected x-ray absorption spectroscopy 
revealed a spectral change 
at low temperatures. 
Using density functional theory (DFT) simulations, we find that
the temperature-induced change originates from a structural phase transition, 
similar to the pressure-induced transition in LaO$_{0.5}$F$_{0.5}$BiS$_2$.
This finding suggests that the mechanism of bulk superconductivity induced by 
Pb substitution is the same as that under high pressure. 
Furthermore,
a novel low-valence state with a mixture of divalent and trivalent Bi ions
is discovered using hard x-ray photoemission spectroscopy with the aid of DFT calculations.

\end{abstract}


%74.25.Jb: Electronic structure (photoemission, etc.) in "74.00.00 Superconductivity"
%74.62.Dh Effects of crystal defects, doping and substitution in "74.00.00 Superconductivity"
%61.10.Ht X-ray absorption spectroscopy: EXAFS, NEXAFS, XANES, etc. in "61.00.00 Structure of solids and liquids; crystallography"
%74.70.Dd Ternary, quaternary, and multinary compounds (including Chevrel phases, borocarbides, etc.)

\maketitle

%\twocolumn
%\narrowtext
%
%%%%%%%%%%%%%%%%%*main text*%%%%%%%%%%%%%%%%%%%
%
%*******************Introduction

\section{Introduction}

More than 20 years have passed since the beginning of the 21$^{\rm st}$ century,
during which several superconductors have been discovered~\cite{Huebener21}. 
In this period, the concept that the phase of 
the wavefunction governs some physical properties led to a whole 
new category of materials called topological materials~\cite{Hasan10}. 
These two fields  converged in the search for topological superconductors. 
Then, the proposed candidate materials such as Cu$_x$Bi$_2$Se$_3$  initiated a new trend in superconductor research~\cite{Hor10,Qi11}.
Alongside this trend, another Bi-based superconductor,
LaO$_{1-x}$F$_x$BiS$_2$, was discovered in 2012~\cite{Mizuguchi12}.
It has a stacking structure consisting of two-dimensional superconducting planes 
and blocking layers, similar to unconventional superconductors 
such as cuprates discovered in the 20$^{\rm th}$ century and later iron-based superconductors~\cite{Stewart11}. 
Unlike them, however, the material has a pyramidal coordination structure with  $C_{4v}$ 
symmetry group at the Bi site, breaking the local inversion symmetry. 
In addition, the 6$p$ electrons in the outermost shell of the Bi atom, 
which are responsible for the superconductivity in this system~\cite{Usui12}, 
exhibit a spin-orbit interaction that is
an order of magnitude stronger than that of the 3$d$ electrons in Cu and Fe. 
The strong spin-orbit interaction induces a large Rashba splitting 
owing to the spatial inversion symmetry breaking~\cite{Wu17}. 
This spin-dependent splitting is, however, 
compensated for by 
the upper and lower BiS layers, resulting in  spin-degenerate bulk bands 
that mask the unique nature of the system~\cite{Zhang14}.

Nevertheless, an unusual property, {\it i.e.} anisotropic upper critical field 
beyond the Pauli limit, has recently been  
reported in the BiS$_2$-based superconductors~\cite{Chan18,Hoshi22}. 
Together with other experimental results suggesting 
unconventional superconductivity and a new pairing mechanism~\cite{Ota17,Hoshi18}, 
this has generated considerable interest in the field.
While the family of BiS$_2$-based superconductors  has expanded 
as a result of extensive research, 
a systematic understanding linking the following points has not yet been fully 
established~\cite{Mizuguchi19, Suzuki19}:
(i) changes in transition temperatures and superconducting volume fractions 
due to pressure application and element substitution,
(ii) microscopic understanding of the flatness of the BiS plane 
and the symmetry of the Bi sites, and
(iii) whether the superconducting mechanism is conventional or unconventional.

In LaO$_{0.5}$F$_{0.5}$BiS$_2$, electron doping of the BiS planes 
by F substitutions on the O sites of the semiconducting LaOBiS$_2$ 
induces superconductivity with a critical temperature $T_{\rm c}\simeq$3~K~\cite{Mizuguchi19}. 
The resulting superconductivity is known to be filamentary with a very low superconducting volume fraction. 
Similar to many other BiS$_2$-based superconductors, this material crystallizes 
in a tetragonal crystal belonging to the space group $P4/nmm$. 
Bulk superconductivity can be observed by substitution on chalcogen and rare-earth sites, 
but there is no significant increase in $T_{\rm c}$~\cite{Demura13,Jha13,Tanaka15,Nagasaka17}.
On the other hand, under high pressure, it exhibits bulk superconductivity with
the highest $T_{\rm c}$ of 11~K in the family, and 
the crystal structure changes to an orthorhombic crystal belonging to the space group $P2_1/m$~\cite{Tomita14}. 
Previous studies have suggested that this temperature-induced structural phase 
transition may even change the mechanism of superconductivity~\cite{Yamashita21}. 
The presence of two adjacent superconducting phases, conventional and unconventional, 
makes it difficult to understand the superconductivity of this system.

Recently, it has been reported that bulk superconductivity appears 
at relatively low Pb concentrations (8-9\%)
after the disappearance of filamentary superconductivity 
when the key element for superconductivity, Bi, is 
partially replaced by Pb~\cite{Otsuki18}. 
It is still unclear whether the superconductivity induced by the 
elemental substitution at the Bi site differs from that in other 
elemental substitution systems. 
In the narrow concentration range of Pb, 
where bulk superconductivity appears, a cusp-shaped electrical 
resistivity anomaly is observed around 100 K, which is not found in other BiS$_2$ systems. 
This anomaly is of interest in the context of superconductivity.

In this study, we investigate the bulk electronic structure of 
LaO$_{0.5}$F$_{0.5}$Bi$_{1-x}$Pb$_x$S$_2$ single crystals with the above Pb concentration
using two types of hard x-ray spectroscopy: x-ray absorption spectroscopy 
and photoemission spectroscopy. 
Our goal is to gain insight into the bulk superconductivity 
and the anomaly around 100 K induced by Pb substitution. 
To reveal subtle temperature changes in the electronic structures, 
we use a high-energy-resolution fluorescence detection mode for 
x-ray absorption near edge structure (HERFD-XANES), 
which can suppress the lifetime broadening effect in the spectrum.
Density functional theory (DFT) simulations show that the spectral change is 
explained by a temperature-induced structural phase transition,
suggesting that the bulk superconductivity is driven by a  mechanism  
similar to that in the high-pressure phase  of LaO$_{0.5}$F$_{0.5}$BiS$_2$.
Hard-x-ray photoemission spectroscopy (HAXPES) results for the Bi valence
are in good agreement  
with the DFT results, indicating the observation of 
a low-valence state with a mixture of divalent and trivalent Bi ions.

%*******************experimental
\section{experimental}
%%%%%%%%%%%%%%%%%%%%%%%%%%%%%%%%%%%

Single crystals of LaO$_{0.5}$F$_{0.5}$Bi$_{1-x}$Pb$_x$S$_2$
($x$=0, 0.05, 0.09) and  LaOBiS$_2$
were grown by a flux method using CsCl and KCl as flux materials, 
where $x$ is the nominal content.
Filamentary and bulk superconductivity appeared at 3~K for $x$=0 and 4~K for $x$=0.09, respectively,
while no superconductivity above 2~K 
was observed  for $x$=0.05.
Details of the sample preparation and their properties have been reported 
elsewhere~\cite{Otsuki18}.
In these compounds,  the typical crystal size was about 1.5~mm$\times$1.5~mm$\times$0.1~mm. 
The crystal structures were determined by powder x-ray diffraction (XRD) 
at the beamline BL02B2 of SPring-8. 
In the case of $x$=0 (0.08), the lattice constants $a$ 
and $c$, analyzed by the Rietveld method
based on the  space group $P4$/{\it nmm}, 
were 4.06263~{\AA} and 13.51420~{\AA}
 (4.06451~{\AA} and 13.47650~{\AA}), respectively~\cite{Demuraxx}. 
Because of the narrow range of Pb concentration at which the compound exhibits bulk superconductivity, 
crystals grown even in the same batch may have different characteristics.
In this study, crystals with $x$=0.09 were selected  for spectroscopic experiments 
on the basis of  the (004) peak position of the laboratory-XRD profile, which is closely related to 
the value of the diamagnetic susceptibility at low temperatures~\cite{Otsuki18}.

The HERFD-XANES experiments were performed at the beamline BL39XU
of SPring-8~\cite{Kawamura17}.
The incident x-ray beam was delivered by the planar undulator and 
monochromized by the Si (220) double-crystal monochromator.
The fluorescence intensity of  Bi and Pb $L\alpha_1$ main lines, selected by 
the Si (911) crystal-analyzer spectrometer, was monitored 
using the two-dimensional pixel detector PILATUS
while sweeping the incident photon energy around  Bi and Pb $L_3$ absorption edges.
The $L\alpha_1$ fluorescence detection overcomes the common 
$L_3$-edge lifetime broadening ($\sim$5.9~eV)
known as the natural linewidth and provides a spectrum with higher 
energy resolution~\cite{Kawamura17,Hamalainen91}.
The energy resolution estimated from an elastic scattering peak near the Bi $L\alpha_1$ line
was 1.1~eV
as the full width at  half maximum (FWHM),
while the lifetime broadening is about 2.9~eV.
In addition, the conventional  
$L_3$-edge XANES spectrum
was  simultaneously recorded by the
partial fluorescence yield (PFY) mode using 
a silicon drift detector.

The HAXPES experiments were performed at the  beamline BL19LXU of 
SPring-8~\cite{Yabashi_01}.
The linearly polarized x-ray was delivered by the 27 m-long undulator, and 
monochromized by the Si (111) double-crystal and Si (620) channel-cut monochromators.
The polarization direction was set to be parallel to the photoelectron scattering plane 
(referred to as $p$ polarization)~\cite{Fujiwara16}.
The incident photon energy and total energy resolution were set to
7.9~keV and 300~meV (FWHM), respectively.
Along the entrance slit,
photoelectrons emitted within approximately $\pm10$ deg of normal to the sample surface 
were detected by the MB Scientific A-1~HE spectrometer
using the A1L2 lens in HE12 mode~\cite{note_e2}.
Clean (001) surfaces were obtained by cleaving the 
samples {\it in situ} in ultrahigh vacuum
($\lesssim1\times10^{-7}$ Pa).
The Fermi energy ($E_{\rm F}$) was determined from the photoemission spectra of evaporated gold films.

To reproduce the spectra, 
we performed scalar-relativistic 
DFT calculations using the {\footnotesize{WIEN2k}} 
code~\cite{Blaha01}.
The local spin density approximation plus spin-orbit coupling was used in the calculations.
For the DFT calculations of
LaO$_{0.5}$F$_{0.5}$BiS$_2$ in the high-pressure (HP) phase and LaOBiS$_2$,
the  lattice constants and atomic coordinates have been taken from Refs.~\onlinecite{Sagayama15,Tomita14}.

%%%%%%%%%%%%%%%%%%%%%%%%%%%%%%%%%%%
%%%%%%%%

%%%%%%%%
\begin{figure}
\includegraphics[width=7.5cm,clip]{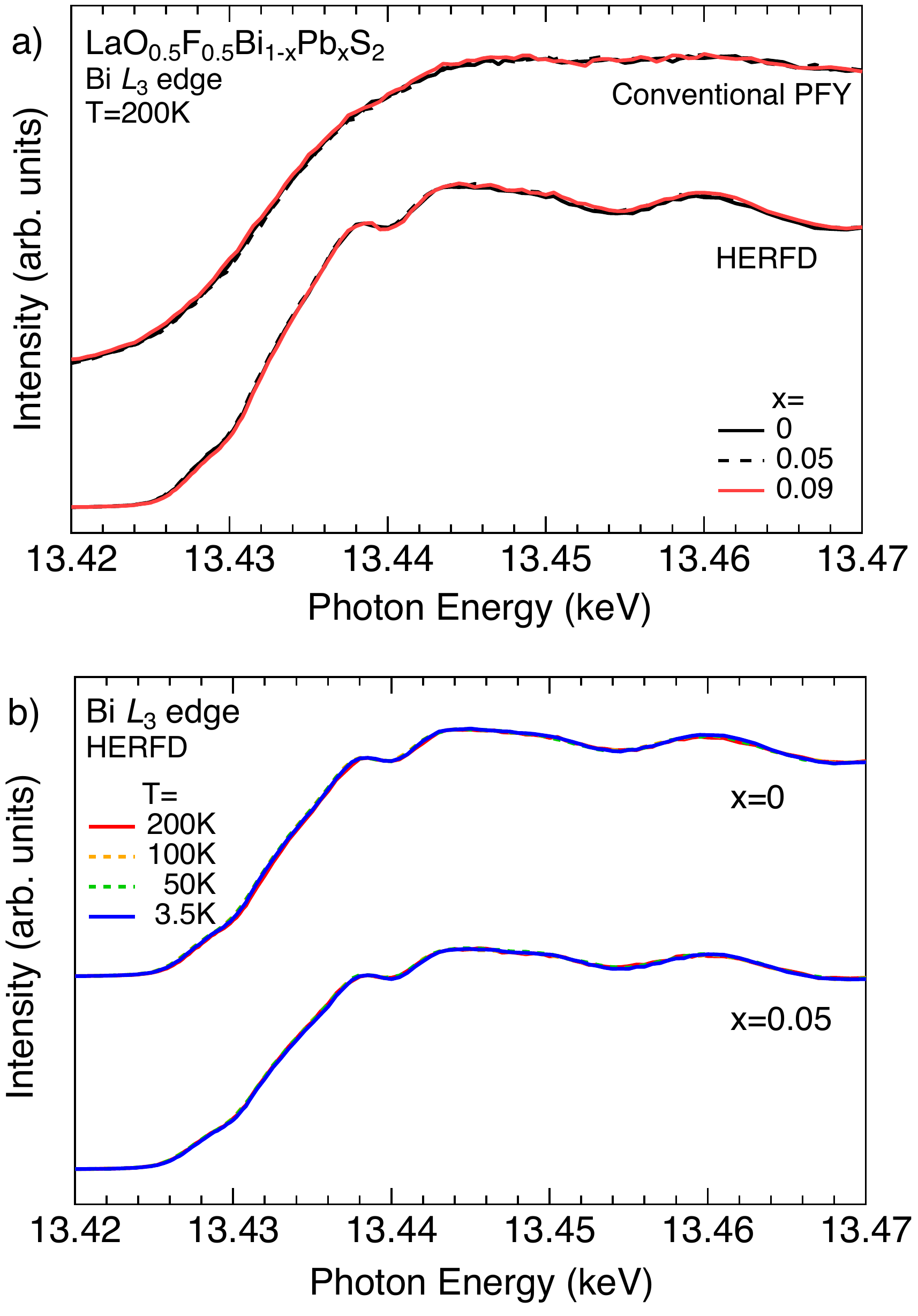}
\caption{
(a)  Bi $L\alpha_1$ HERFD-XANES spectra for $x$=0, 0.05, and 0.09 at 200~K.
Simultaneously measured conventional 
PFY spectra are also displayed.
(b)Temperature dependence of Bi  $L\alpha_1$ HERFD-XANES spectra for $x$=0 and 0.05.
}
\label{Fig_1}
\end{figure}
%%%%%%%%%%%%%%%%%%%%%%%%%%%%%%%%%%%
%%%%%%%%
\begin{figure}
\includegraphics[width=8cm,clip]{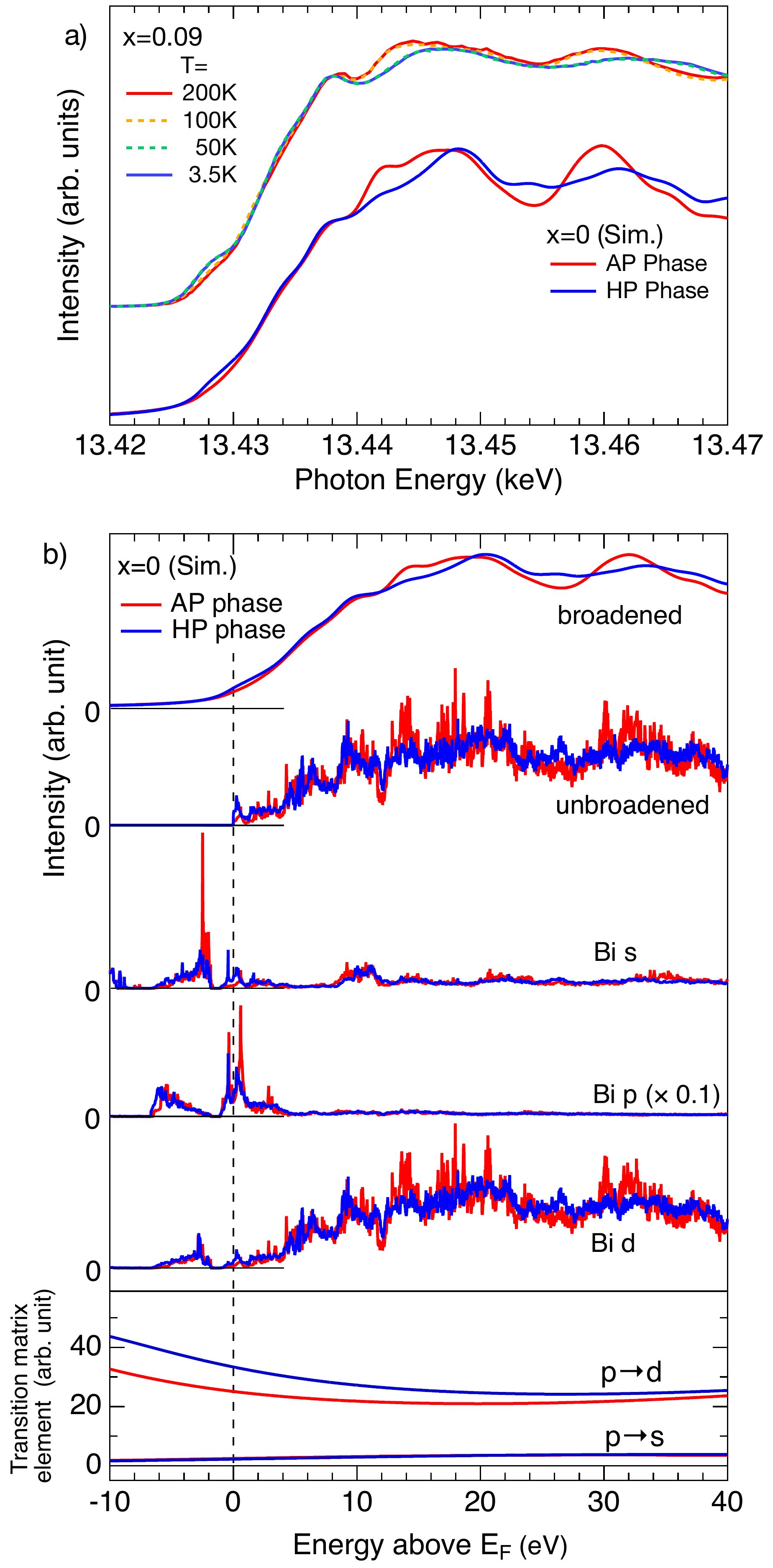}
\caption{
Experimental and simulated Bi $L_3$-edge XANES spectra of LaO$_{0.5}$F$_{0.5}$Bi$_{1-x}$Pb$_x$S$_2$.
(a) Temperature dependence of Bi $L\alpha_1$ HERFD-XANES spectrum  for $x$=0.09.
Simulated spectra for $x$=0 in the AP and HP phases are shown for comparison.
The simulated spectra are broadened by Gaussian and Lorentzian functions 
representing the experimental energy-resolution and lifetime  effects.
(b) Details of the simulated spectra in (a). These are composed of 
unoccupied Bi  $s$- and $d$-orbital pDOSs  after
multiplying the transition probabilities.
The Bi $p$-orbital pDOS is also shown for reference.
}
\label{Fig_2}
\end{figure}
%%%%%%%%%%%%%%%%%%%%%%%%%%%%%%%%%%%

%*******************results
\section{results and discussion}
%%%%%%%%%%%%%%%%%%%%%%%%%%%%%%%%%%%
\subsection{HERFD XANES}

Figure \ref{Fig_1}(a) shows  Bi $L_3$-edge XANES spectra of LaO$_{0.5}$F$_{0.5}$Bi$_{1-x}$Pb$_x$S$_2$
for $x$=0, 0.05, and 0.09 recorded at $200$~K using two different techniques, {\it i.e.} 
the conventional PFY  and the HERFD modes.
In contrast to the featureless and rather broadened PFY spectra,
some fine structures  are resolved in the HERFD spectra.
Meanwhile, the spectra for $x$=0, 0.05, and 0.09 are barely distinguishable from one another 
in both modes.
The temperature dependence of the HERFD-XANES spectra for $x$=0 and 0.05 is shown in Fig.~\ref{Fig_1}(b).
The spectra do not show any temperature dependence.
In contrast, Fig.~\ref{Fig_2}(a) shows that the spectral shape at $x$=0.09 changes significantly between 100 and 50~K.
The temperature range in which the spectral change was observed 
includes the temperature at which 
the cusp-shaped anomaly in the $\rho-T$ curve was observed~\cite{Otsuki18}.
In addition, a synchrotron-XRD experiment has recently demonstrated 
the peak splitting in the profile for $x$=0.09 below about
100~K~\cite{Demuraxx}, suggesting the structural transition of the crystal with the symmetry lowering.

%%%%%%%%
\begin{figure}
\includegraphics[width=8cm,clip]{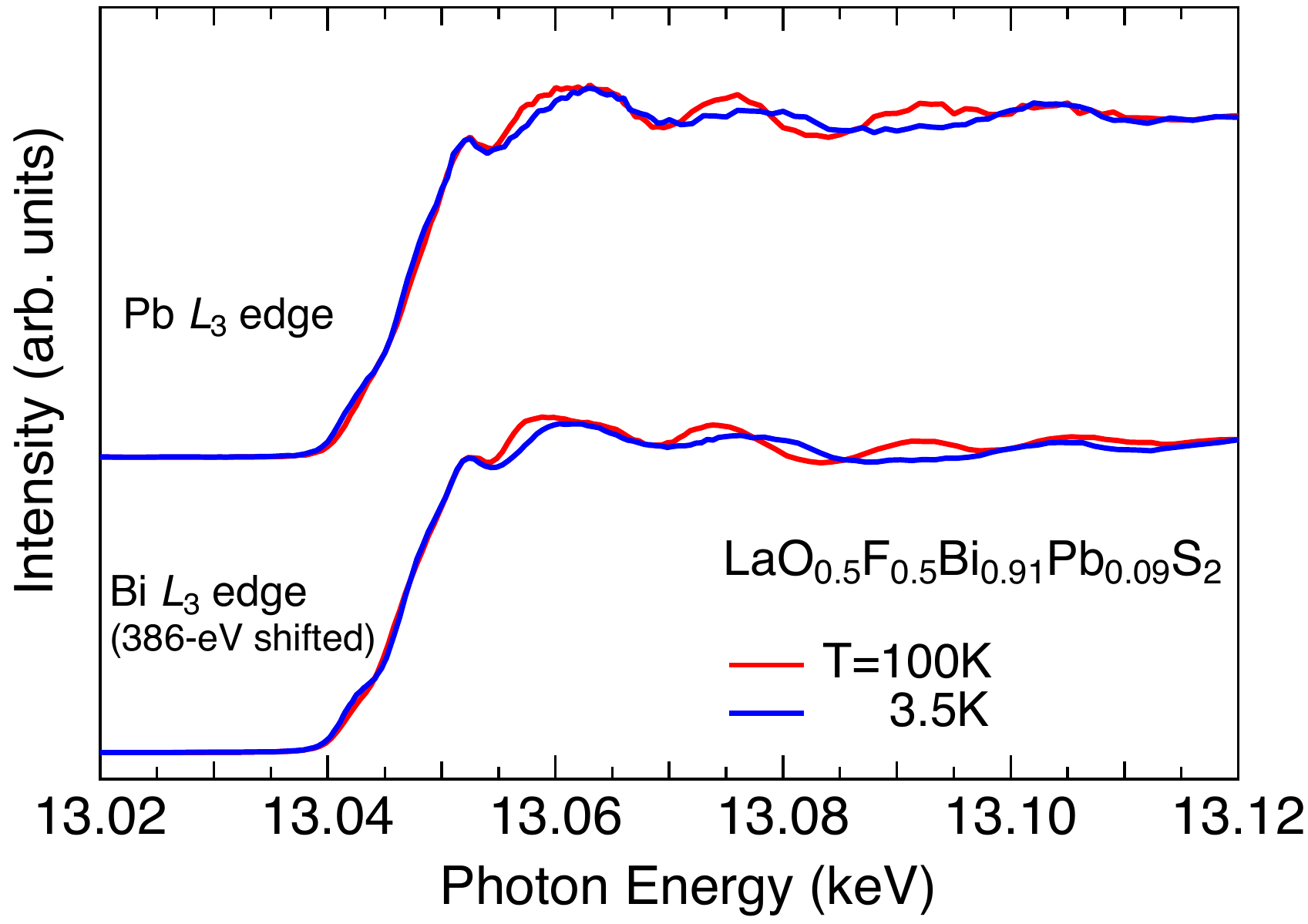}
\caption{
Temperature dependence of  Pb  $L\alpha_1$ HERFD-XANES spectrum  for $x$=0.09.
Bi $L\alpha_1$ HERFD-XANES spectra (shifted)  are also shown for comparison.
}
\label{Fig_3}
\end{figure}
%%%%%%%%%%%%%%%%%%%%%%%%%%%%%%%%%%%

To reproduce the temperature dependence of the HERFD-XANES spectrum for $x$=0.09, 
we  performed  DFT calculations and
simulated the spectra using the {\footnotesize{XSPEC}} program~\cite{Blaha20,DFTconditions}.
Since  details of the crystal structure of the low-$T$ phase for $x$=0.09 are not yet known,
we  assumed here that the phase has the same crystal structure,
{\it i.e.} the same lattice constants and atomic coordinates, as 
the HP phase for $x$=0 with the space group $P2_1/m$,
based on the experimental evidence: high superconducting volume fraction and 
 high $T_{\rm c}$ in both phases~\cite{Tomita14}.
In addition, since the spectral shapes of $x$=0 and 0.09 at high temperatures are very similar
as seen in Fig.~\ref{Fig_1}(a),
the crystal structure of $x$=0 in the ambient-pressure (AP) phase is used for comparison
instead of that of $x$=0.09.
The simulated XANES spectra are shown in Figs.~\ref{Fig_2}(a) and (b).
The total difference induced by the temperature change 
is well reproduced by the simulation,
indicating that the temperature-induced change originates from a structural phase transition, 
similar to the pressure-induced transition in LaO$_{0.5}$F$_{0.5}$BiS$_2$.
This suggests that the mechanism of bulk superconductivity induced by 
Pb substitution is the same as that under high pressure. 
In the Bi $L_3$-edge absorption process, the electrical dipole transition from $2p$ to $6d$ states becomes dominant,
as can be seen at the bottom of Fig.~\ref{Fig_2}(b).
Therefore, the XANES spectra mainly reflect unoccupied Bi $6d$ states
depending on their crystal structures,
although one can also find a structural-induced change at the absorption threshold in Bi $6s$ states.

We also investigated the Pb-derived electronic states.
Figure~\ref{Fig_3} shows the temperature variation of the Pb $L\alpha_1$ HERFD-XANES spectrum  for $x$=0.09.
The spectral shape and the temperature-induced change observed here are very similar to those seen 
in the spectrum for the Bi $L_3$ edge, indicating that 
substituted Pb atoms properly occupy the Bi lattice site.

%%%%%%%%%%%%%%%%%%%%%%%%%%%%%%%%%%%
\begin{table*}
\caption{
Bi and Pb valences in the BiS$_2$-based compounds.
The nominal Bi valence is determined by assuming
La$^{3+}$, O$^{2-}$, F$^{-}$, S$^{2-}$, and Pb$^{2+}$ ions.
Experimental values are estimated from Bi and Pb $4f_{7/2}$ core-level HAXPES spectra.
Errors are attributed to the uncertainty of possible background.
DFT values are derived from  $s$-, $p$-, and $d$-orbital pDOSs integrated up to $E_{\rm F}$  in the atomic sphere.
}
\begin{ruledtabular}
\begin{center}
\begin{tabular}{lcccccccc}
Compound & $T$(K) & \multicolumn{3}{c}{Bi valence}  &{} & \multicolumn{3}{c}{Pb valence} \\ \cline{3-5} \cline{7-9}
{} & {} & nominal & HAXPES & DFT&{}   & nominal & HAXPES & DFT \\  \hline
LaOBiS$_2$ & 50 &  3 &  2.82(4) & 2.70\\
LaO$_{0.5}$F$_{0.5}$Bi$_{1-x}$Pb$_x$S$_2$  & & & & \\
\ \ \ \ \ $x$=0        & 50 & 2.5   & 2.54(4) & 2.65\\
\ \ \ \ \ $x$=0.09     & 200 & 2.55  &  2.61(2) & - & {} & 2 & 2.48(2) & -\\
             & 50 & 2.55  & 2.63(3) & - & {} & 2 & 2.48(2) & -\\
\ \ \ \ \ $x$=0.125    & high-$T$ phase& 2.57  & - & 2.68  & {} & 2 & - & 2.39\\
\ \ \ \ \ \ \ \ \ \ \ \ \ \ \ \ \      & low-$T$ phase&  2.57  &  - & 2.63 & {} & 2 & - & 2.33\\
\end{tabular}
\end{center}
\end{ruledtabular}
\label{Table_1}
\end{table*}

%%%%%%%%%%%%%%%%%%%%%%%%%%%%%%%%%%%

%%%%%%%%
\begin{figure*}
\includegraphics[width=12.0cm,clip]{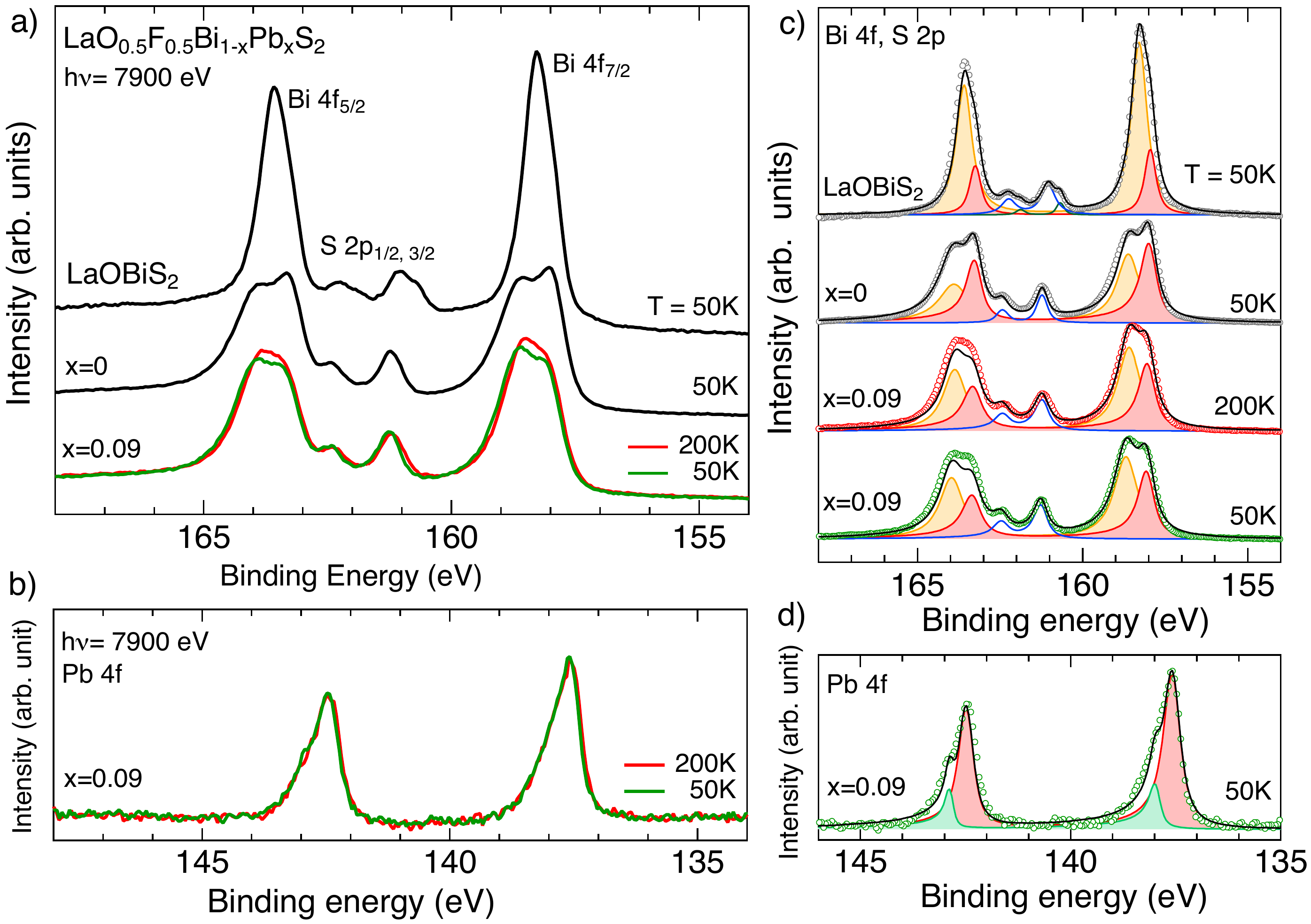}
\caption{
Bi and Pb $4f$ core-level HAXPES spectra of LaO$_{0.5}$F$_{0.5}$Bi$_{1-x}$Pb$_x$S$_2$.
The spectra are normalized by the area under the curves after subtraction of the Shirley-type background~\cite{Shirley72}.
(a) Bi $4f$ core-level HAXPES spectra, including the S $2p$ core levels, for $x$=0 and 0.09.
The spectrum of LaOBiS$_2$ is also shown.
(b) The temperature dependence of the Pb $4f$ core-level HAXPES spectrum  for $x$=0.09.
(c) and (d) The results of the line-shape analysis for the HAXPES spectra.
}
\label{Fig_4}
\end{figure*}
%%%%%%%%%%%%%%%%%%%%%%%%%%%%%%%%%%%

\subsection{Core-level HAXPES}

The Bi- and Pb-derived  electronic states  were further investigated by  
HAXPES, which is a complementary method for probing  bulk electronic structures.
Bi $4f$ core-level HAXPES spectra of LaOBiS$_2$ and 
LaO$_{0.5}$F$_{0.5}$Bi$_{1-x}$Pb$_x$S$_2$ for $x$=0 and 0.09 are shown in Fig.~\ref{Fig_4}(a).
LaOBiS$_2$ has a single-peak-like structure in both the Bi 4$f_{7/2}$ 
and 4$f_{5/2}$ states,
as  observed in the soft-x-ray (SX) photoemission study~\cite{Nagira14}.
In contrast, a double-peak structure with  stronger intensity of the lower-binding energy ($E_{\rm B}$) peak 
was observed  for $x$=0.
This characteristic feature in the HAXPES spectrum is quite different from that seen in the SXPES spectrum,
which has the single-peak structure with a weak shoulder on the lower-$E_{\rm B}$ side.
Such a difference is often observed in some correlated materials
where the electron correlation effects are enhanced at the surface.
Owing to the long inelastic mean free path of emitted photoelectrons~\cite{Tanuma11},
HAXPES provides highly bulk sensitive information, 
resulting in a significant intensity of the non-locally screened {\it final} state 
in the spectrum for these materials~\cite{Taguchi05}.
However, it should be noted that BiS$_2$-based compounds
are considered to be an itinerant electron system,
as the highly dispersive Bi $6p$ bands  cross the $E_{\rm F}$ in the metallic state and 
play a key role in the low-energy phenomena such as
superconductivity~\cite{Usui12}.
In this case, it is reasonable to assume that the double-peak structure observed 
in the Bi $4f$ core level consists of two states with different valences in the 
{\it initial} state, as in other Bi compounds~\cite{Chaturvedi14,Nishikubo19}.
Thus, the spectral change resulting from the F (Pb) substitution to the O (Bi) site can be  understood 
in terms of the electron (hole) doping effect.
As for $x$=0.09, the lower-$E_{\rm B}$ peak is suppressed by the Pb substitution in the double-peak structure, 
which becomes slightly broader and shifts ($\sim$50~meV) with decreasing temperature.

\begin{figure}
\includegraphics[width=7cm,clip]{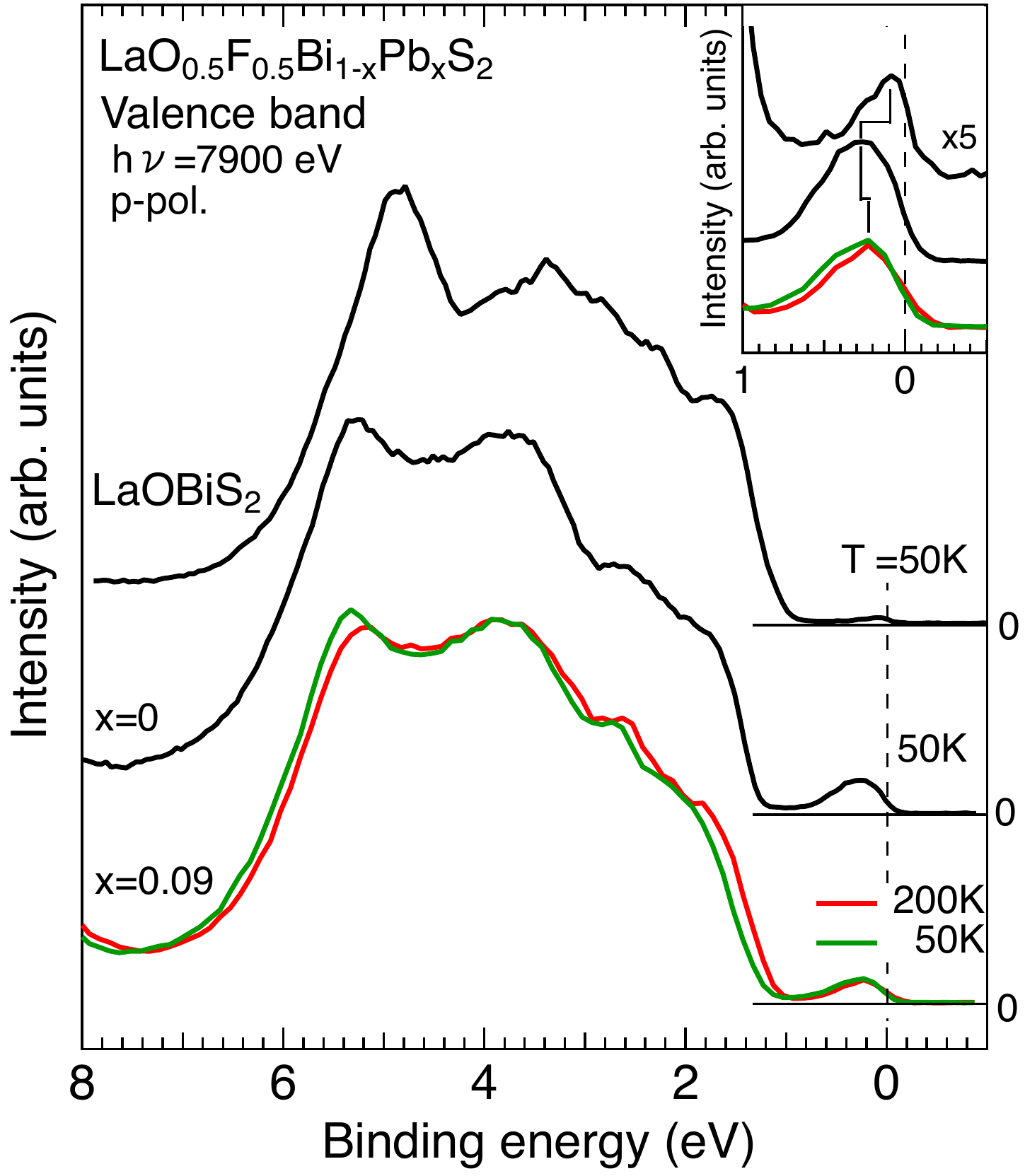}
\caption{
VB-HAXPES spectra of LaO$_{0.5}$F$_{0.5}$Bi$_{1-x}$Pb$_x$S$_2$.
The spectrum of LaOBiS$_2$ is also shown.
The inset shows the enlarged spectra near $E_{\rm F}$.
The spectra are normalized by the area under the curves after subtraction of the Shirley-type background.
}
\label{Fig_5}
\end{figure}

Line-shape analysis shows that the spectral shape of each of the Bi 4$f_{7/2}$ or 4$f_{5/2}$ state
in all three compounds can be accurately 
reproduced by two Voigtian peaks, as shown in Fig.~\ref{Fig_4}(c).
Although  Bi ions tend to be  neutral, trivalent, and pentavalent,
the lower-$E_{\rm B}$ (higher-$E_{\rm B}$) component is assumed to be derived from the divalent (trivalent) state
by considering the following facts: 
(i) the nominal valence of Bi ions for semiconducting LaOBiS$_2$ is +3,
(ii) the valence-band (VB) HAXPES spectrum of LaOBiS$_2$ shown in the inset of Fig.~\ref{Fig_5}
indicates  that
a small amount of  electrons are doped to Bi 6$p$ bands possibly owing to some oxygen 
and/or sulfur defects, 
yielding  lower-valence state.
In addition, (iii)  electron doping  by F substitution leads to an increase in the lower-$E_{\rm B}$  peak;
(iv) the lower-$E_{\rm B}$  peak decreases with hole doping owing to Pb substitution in the F-substituted system.
The Bi valences estimated by the line-shape analysis are listed in Table~\ref{Table_1}.
Furthermore, Fig.~\ref{Fig_4} (b) shows  Pb $4f$ core-level HAXPES spectra obtained at 200 and 50~K.
In contrast to the  Bi $4f$ core level, the spectrum  has a single peak with strong asymmetry, in which 
the temperature-induced change is hardly visible except for the slight shift ($\sim$30~meV).
This  shift would originate from the structural transition which induces the change in the potential at the Pb site. 
The Pb valences estimated by the line-shape analysis (shown in Fig.~\ref{Fig_4} (d)) are also listed in Table~\ref{Table_1}.
They will be discussed later together with the DFT results.

%%%%%%%%
%%%%%%%%%%%%%%%%%%%%%%%%%%%%%%%%%%%

\subsection{Valence-band HAXPES}

\begin{figure}
\includegraphics[width=8.6cm,clip]{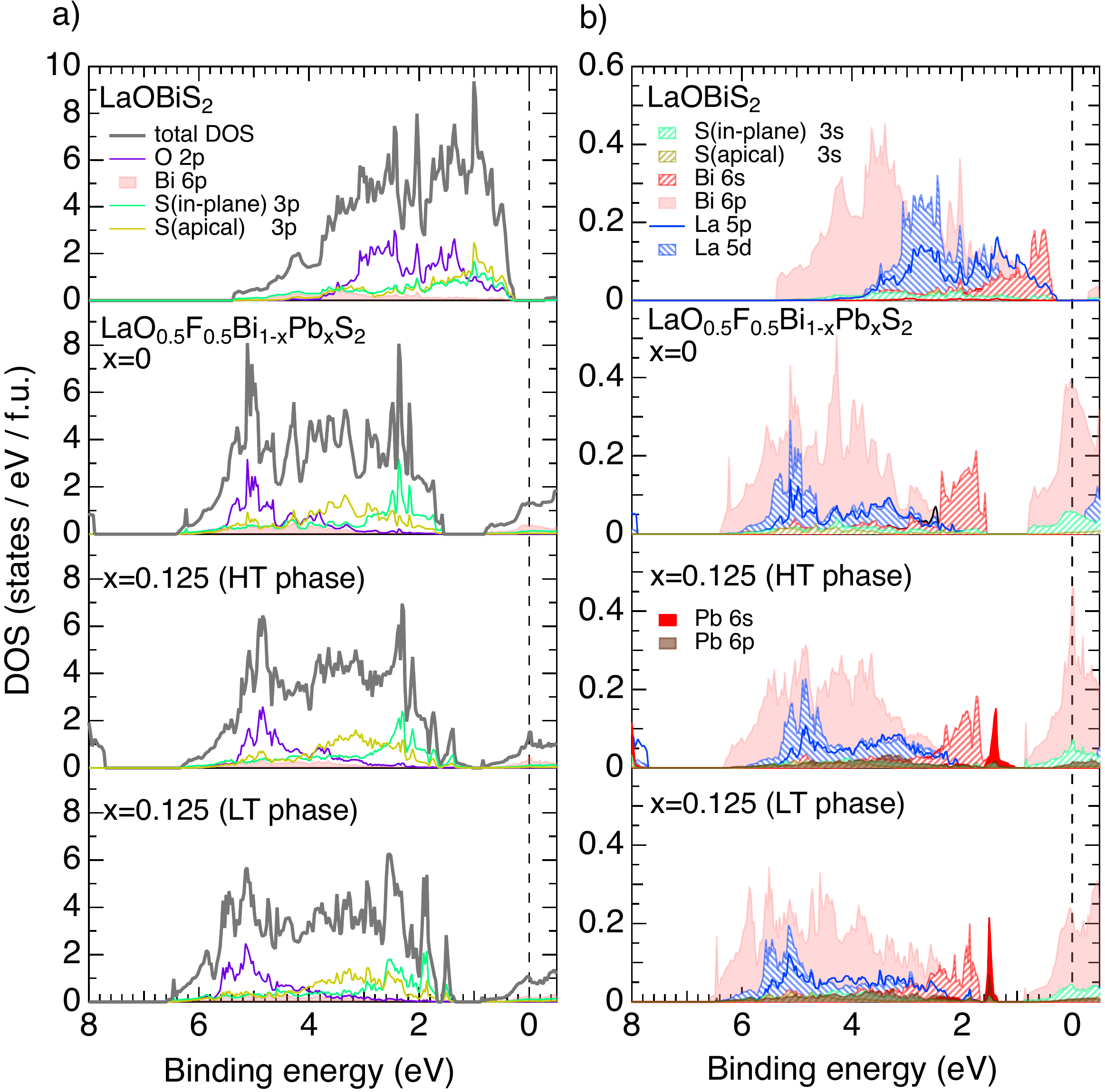}
\caption{
Total DOS and some pDOSs of LaOBiS$_2$ and LaO$_{0.5}$F$_{0.5}$Bi$_{1-x}$Pb$_x$S$_2$.
(a) Overall and (b) enlarged views.
}
\label{Fig_6}
\end{figure}

To investigate the effects of element substitution and temperature variation on the electronic structures near $E_{\rm F}$,
we measured  VB-HAXPES spectra.
Figure~\ref{Fig_5} shows the spectra obtained with the $p$-polarized HAX.
Compared to the SXPES~\cite{Nagira14}, which is sensitive to S $3p$ states, 
the HAXPES shows an increase in intensity around 3-6 eV.
For example, in the case of  LaO$_{0.5}$F$_{0.5}$BiS$_2$,
DFT calculations show that O $2p$ states dominate in this energy range;
S $3p$ and Bi $6p$ bonding states are also present, as shown in Fig.~\ref{Fig_6}. 
The intensity derived from the bonding states is enhanced in the $p$-polarized HAXPES spectrum 
because the photoionization cross section of the Bi $6p$ states is relatively large under the present experimental conditions~\cite{Trzhaskovskaya06}.

As shown in Fig.~\ref{Fig_5}, semiconducting LaOBiS$_2$ has a very weak peak near $E_{\rm F}$, 
indicating that a small amount of electrons are doped into Bi $6p$ dominant bands as mentioned above.
By substituting O for F, the peak develops 
and shifts  to the higher $E_{\rm B}$ side by 180~meV
(see inset for illustration),
indicating  electron doping in these bands.
In contrast, for the Pb substitution,  the peak was observed to 
shift 50~meV toward  $E_{\rm F}$ owing to  hole doping.
These results support the interpretation that the change in the spectral shape of the Bi $4f$ core level 
due to F (Pb) substitution is attributed to electron (hole) doping.

As for the temperature dependence of the spectrum at $x$=0.09, a slight change in the spectral shape was observed, accompanied by a shift of about 110 meV to the high $E_{\rm B}$ side in the low-$T$ phase compared to the high-$T$ phase, rather than a large change in the broad energy range as in the HERFD-XANES spectrum.
This spectral shift is larger than those in the Bi $4f$ and Pb $4f$ core levels (60 meV and 30 meV, respectively) and is thought to reflect a change in the VB electronic structure with symmetry lowering caused by the structural phase transition to the low-$T$ phase.

%%%%%%%%
%%%%%%%%%%%%%%%%%%%%%%%%%%%%%%%%%%%

\subsection{Mixed-valence state}
Next, we  discuss the validity of the experimentally obtained valences, which are summarized in Table~\ref{Table_1}.
To evaluate the Bi and Pb valences, we performed further DFT calculations for all other compounds and phases, whose DOSs are also shown in Fig.~\ref{Fig_6}~\cite{DFTconditions_2}. 
Note that in the calculations, the percentage of Pb substitution was assumed to be 12.5 \% instead of the actual 9\% to save  computational effort for Pb-substituted systems. 
The valences estimated from the  DFT results 
are also listed in Table~\ref{Table_1}.
Theoretical valences are obtained by the sum of integrated
$s$-, $p$-, and $d$-orbital pDOSs up to $E_{\rm F}$ 
within the atomic sphere of 1.32\AA \ for both Bi and Pb atoms,
which is greater than or equal to the ionic radii of octahedrally 
coordinated Bi$^{3+}$, Bi$^{2+}$, and Pb$^{2+}$ ions~\cite{Shannon76,note_r1}.
The experimental value of the Bi valence in LaOBiS$_2$,  which is closest to +3 in the three compounds,
decreases to about +2.5 with 50\% F substitution.
This is in good agreement with the DFT prediction.
While the Bi$^{3+}$ state is the most chemically stable, the Bi$^{2+}$ state has been reported to be stabilized in Bi-doped materials such as Sr$_2$P$_2$O$_7$:Bi$^{2+}$~\cite{Li15}.
In contrast, to the best of our knowledge, mixed valence states of divalent and trivalent Bi  have been found in very few materials such as  Bi-doped zeolites~\cite{Bai12}.
Our results show that
the mixed-valence states are realized in BiS$_2$-based 
compounds, which can lead to some charge-ordered structures 
and/or microscopic charge disproportionation~\cite{Machida14}.
Furthermore, the HAXPES captures the subtle increase in Bi valence caused by 
Pb substitution in the AP ({\it i.e.}, high-$T$) phase, which is also 
reproduced by the DFT calculation.
In contrast,  the valence at 200~K is more or less the same as at 50~K,
although it predicts a slight decrease with the phase transition.
Although the electronic structure of Pb changes below 100~K,
as seen in the Pb $L\alpha_1$ HERFD-XANES spectra,
the valence of Pb does not change, unlike that of Bi.
Under the assumption that the large (small) component in the spectrum shown in Fig.~\ref{Fig_4}(d) is derived from
the Pb$^{2+}$ (Pb$^{4+}$) state, the mean valence is +2.48~\cite{Sakai17}.
On the other hand, it becomes +2.24 if the chemically unstable Pb$^{3+}$ state possibly causes 
the small component.

Note that the above estimates are based on the assumption 
that all collected photoelectrons came from the bulk. 
At the surface, the Bi valence has been reported 
to be closer to +3 than in the bulk~\cite{Terashima16}. 
However, even assuming pure Bi$^{3+}$ state 
at the surface, 
the influence of the surface on the estimates is 
within the errors owing to the high bulk sensitivity of 
HAXPES.

\section{summary}

Taking advantage of the high energy resolution, we found that the HERFD-XANES spectra 
show a clear temperature change only in 
LaO$_{0.5}$F$_{0.5}$Bi$_{1-x}$Pb$_x$S$_2$ ($x$=0.09), 
where bulk superconductivity is observed.
This change is due to the symmetry lowering of the crystal 
at low temperatures, suggesting that the bulk superconductivity in the Pb-substituted BiS$_2$ 
compound is driven by a  mechanism  similar to that in the HP phase  
of LaO$_{0.5}$F$_{0.5}$BiS$_2$.
The HERFD-XANES, core-level HAXPES, and VB-HAXPES spectra are  consistently understood by the
DFT calculations.
With the help of the DFT calculations, Bi 4$f$ core-level HAXPES revealed that 
the doped BiS$_2$-based compounds host a mixed state of divalent and trivalent Bi ions
at the  site of broken local inversion symmetry. 
This is a novel case for Bi compounds.

%%%%%%%%%%%%%%%%%%%%%%%%%%%%%%%%%%%
\begin{acknowledgments} 

We would like to thank Y. Murakami, Y. Arinaga, Y.~Kondo, K. Oda, and M. Shimamoto for supporting the HAXPES experiments,
A. Hariki for fruitful discussion, M. Matsuki-Baltzer for providing 
support information.
The HERFD-XANES and the synchrotron-XRD 
experiments at SPring-8
 were performed  with the approval of the Japan Synchrotron Radiation Research Institute
(Proposal No.~2020A0630 and No.~2020A1475,
respectively);
the HAXPES experiments at SPring-8
were performed  with the approval of
RIKEN (Proposal Nos.~20190031, 20200075, and 20210068), 
under the support of JSPS Grant-in-Aid for Scientific Research (C)(Nos.~19K03753 and 22K03527)
from the Ministry of Education, Culture, Sports, Science, and Technology, Japan.

\end{acknowledgments}

%%%%%%%%%%%%%%%%%%%%%%%%%%%%%%%%%%%%%%%%%%%%%%%

%\appendix*

%\end{multicols}


\begin{references}
\bibitem{DFTconditions}
For the AP (HP) phase,
we used the  {$\boldsymbol k$}-point mesh of $11\times11\times3$ 
with 72 (115) irreducible {$\boldsymbol k$} points
for the $2\times2\times2$ supercell; 
the cutoff parameter, $R_{\rm MT}K_{\rm max}$, was set to 6.5 (7.0) for 
the minimum muffin-tin radius of 2.18 (2.01) bohr.
\bibitem{DFTconditions_2}
For LaOBiS$_2$ and $x$=0,
we used the  {$\boldsymbol k$}-point mesh of $11\times11\times3$
for the unit cell;
for $x$=0.125 in both low-$T$ and high-$T$ phases, 
we used $9\times9\times5$ for the $2\times2\times1$ supercell.
$R_{\rm MT}K_{\rm max}$ was set to 6.0-7.0 for these compounds.
\bibitem{note_r1}
Some properties of  Bi$^{2+}$ ion including the ionic radius have not been widely reported because of  chemical instability. It is reasonable to assume that the radius of  Bi$^{2+}$ ion is smaller than that of  Pb$^{2+}$ ion, considering that the atomic radius of Bi is smaller than that of Pb.
\end{references}
\end{document}